\def\G1915{GRS $1915$+$105$}
\def\X1550{XTE J$1550$-$564$}
\def\J1655{GRO J$1655$-$40$}

\def\etal{{\em et al. } }

\def\dd #1 {{\frac{\partial}{\partial #1}}}

\def\cs2{c_{S}^2}

\def\ltsima{$\; \buildrel < \over \sim \;$}
\def\simlt{\lower.5ex\hbox{\ltsima}}
\def\gtsima{$\;\buildrel>\over\sim\;$}
\def\simgt{\lower.5ex\hbox{\gtsima}}

\def\correc#1{ #1}

\documentclass{aa}
\usepackage{color}
\usepackage{graphicx}

\begin{document}
\title{Accretion-ejection instability and QPO in black hole binaries \\ 
I. Observations}
\author{J. Rodriguez \and P. Varni\`ere \and M. Tagger \and Ph. 
Durouchoux}
\offprints{J. Rodriguez \\
(rodrigue@discovery.saclay.cea.fr)}
\institute{Service d'Astrophysique (CNRS URA 2052), CEA Saclay, 91191 
Gif-sur-Yvette, France}
\titlerunning{AEI \& QPO in BH Binaries}
\authorrunning{Rodriguez, Varni\`ere, Tagger \& Durouchoux}
\date{Received 23 May 2000 / Accepted 25 February 2002}
%
\abstract{
This is the first of two papers in which we address the physics of the
low-frequency Quasi-Periodic Oscillation (QPO) of X-ray binaries, in
particular those hosting a black hole.  We discuss and repeat the recent
analysis and spectral modelling of the micro-quasar \J1655 by Sobczak
\etal (2000, hereafter SMR), and compare it with \G1915; this leads us
to confirm and analyze in more detail the different behavior noted by
SMR, between \J1655 and other sources, when comparing the correlation
between the QPO frequency and the disk inner radius.  In a companion
paper (Varni\`ere {\em et al. }, 2002, hereafter Paper II) we will show that
these opposite behaviors can be explained in the context of the
Accretion-Ejection Instability recently presented by Tagger and Pellat
(1999).\\
We thus propose that the difference between \J1655 and other sources
comes from the fact that in the former, observed in a very high state,
the disk inner radius always stays close to the Last Stable Orbit.  \\
In the course of this analysis, we also indicate interesting differences
between the source properties, when the spectral fits give an
anomalously low inner disk radius.  This might indicate the presence of
a spiral shock or a hot point in the disk.  
}
\maketitle

\section{Introduction}
The observation of Quasi-Periodic Oscillations (QPOs) in X-ray binaries
is widely considered as a key to understanding the physics of the inner
region of accretion disks around compact objects.  Although the origin
of the QPOs is still debated, it is considered that such models should
yield information of prime importance on the physics and the geometry of
the accretion process, in these and other accreting sources.  In
particular, in the last years, many observational results have pointed
to the importance of a low-frequency QPO, found in binaries hosting
neutron stars or black holes.  A recent result, reported by Psaltis {\em
et al.} (1999), shows that in these sources its frequency is correlated
with that of a higher-frequency QPO (in neutron-star binaries, this is
the lower of the pair of so-called `kHz QPOs'), believed to be close to
the rotation frequency at the inner edge of the disk.  Thus the
frequency of this QPO ranges from $\simlt$ 1 to a few tens of Hz in
black-hole and neutron-star binaries.  It has been intensively studied,
in particular, in the micro-quasar \G1915, where its occurrence 
 in the low-hard state is so
frequent that it has been dubbed the ``ubiquitous'' QPO (Swank {\em et
al. }, 1997; Markwardt {\it et al. }, 1999, hereafter SM97 for both these
references).  These studies have allowed SM97 and Muno \etal (1999) to
find a number of correlations, showing that the QPO probably has its
origin in the disk, although it affects more strongly the coronal
emission.

One of these correlations involves the QPO frequency and the color
radius of the disk $r_{col}$, considered as a measure of the disk inner
radius $r_{int}$.  Even though this is not a very tight correlation, it
would mean that the QPO frequency would most often correspond to a 
Keplerian rotation frequency in the disk, at a radius of a few
times (typically $\sim 2$ to $\sim 5$) $r_{int}$; this is
consistent with the correlation of Psaltis {\em et al. } (1999), 
where the high-frequency to low-frequency ratio is about 11, assuming, as 
usually believed, that the high-frequency QPO is close to the Keplerian 
frequency at the inner edge of the disk. 

However these observational results must be taken with extreme caution:
$r_{col}$ is obtained from a fit of the source spectrum with a model,
and the results of the fit raise many questions:
\begin{enumerate}
    \item The model spectrum consists of a multi-color black-body part,
    thought to originate in an optically thick disk, and a power-law
    tail at higher energies, considered as inverse Compton emission from
    a hot corona which might lie either above the disk, or in the region
    between the disk and the central object.  We will call them, for
    simplicity, the disk and coronal emissions respectively.  However in
    the low-hard state where the low-frequency QPO is most often seen,
    the coronal emission often dominates the disk one, even at low
    energies.  In this case the extraction of the disk parameters
    becomes difficult.  

    \item Indeed the fits often give values of
    $r_{col}$ which must clearly be ruled out, since they are extremely
    low and sometimes even within the Schwarzschild radius.  

    \item In a recent paper Merloni {\em et al.} (2000) (hereafter MFR)
    have shown that other effects, associated with the vertical
    structure of the disk, could strongly affect the determination of
    the disk parameters.  Using synthetic spectra to test the fitting
    algorithm, they find that the disk black body model underestimates
    the radius by a factor which may be quite large, when the coronal
    emission dominates the disk emission.  This makes the extraction of
    disk parameters difficult, since we have little information on the
    conditions causing this effect.

    \item Even when the disk emission is strong enough to allow a
    reliable fit, the model spectrum contains assumptions which may
    affect the outcome.  In particular the disk is assumed to be
    axisymmetric with a temperature varying as $T^{-3/4}$, following the
    standard $\alpha$-disk model.  These are strong assumptions which
    may affect the validity of the fit parameters.  However, since
    the  RXTE/PCA data starts above 2 keV, the observations are not very 
    sensitive to the whole structure of the disk itself. On the other hand, 
    various corrections ({\em e.g.} due to electron scattering) must be
    applied.  One can only expect that, when the assumptions are not too
    far from reality, there is a proportionality between the color
    radius $r_{col}$ and the disk inner radius $r_{int}$.
\end{enumerate}

Points (1-3) above thus lead us to take with extreme caution, or rule out,
determinations of the disk parameters when the ratio of the disk black
body flux (determined from the fit) is less than about half of the total
flux.  We will see below that indeed, among the data points we 
obtain for \J1655, {\em all} the ones giving anomalously low values of
$r_{col}$ are  excluded if this criterion is applied.
 On the other hand point (4) above is much more
difficult to address, as long as we have no observational tests of the
disk structure.  Determining the nature of the QPO would be very likely
to shed light on this, but this also depends on our knowledge of the
disk parameters.

Recently SMR have analyzed observations of the micro-quasar \J1655, and
found that in this source the low-frequency QPO differed markedly from
other sources, in particular \X1550: the color radius was found to be
quite small (down to a few kms), and to increase with the QPO 
frequency. 
On the contrary in \X1550, and in most observations of other sources,
the color radius and the QPO frequency are inversely correlated.

In this paper our goal is to check this reversed correlation which (as
will be seen in Paper II) might be consistent with the explanation of
this QPO by the Accretion-Ejection Instability (AEI) recently presented
in Tagger \& Pellat (1999).  It would be due to relativistic effects,
when the disk inner radius approaches the Last Stable Orbit.  Thus we
begin in section \ref{subsec:J1655} by reconsidering the data points
from \J1655 presented by SMR, repeating the whole data reduction and
fitting process; we discuss them in regard to the criterion of MFR,
which we take as a guide for the validity of the fitting procedure. 
This results in excluding all the points at anomalously low radius
values, and retaining points only in a limited radial range.  We
then turn to the other source studied by SMR, \X1550, but the criterion
of MFR leads us to exclude almost the whole data set: the source was at 
the time of the study in an extremely bright high soft state/ very high 
outburst
state.  We thus chose to study a different source, \G1915, because its
high variability and the large number of observations including the low
frequency QPO suggested that we could explore the radius-frequency
correlation over a much broader range of values for $r_{col}$.  In
particular the observations of SM97 showed that, during the much studied
30 min.  cycle of that source, both $\nu_{QPO}$ (the QPO frequency) and
$r_{col}$ vary by a large factor, and apparently in opposite directions. 
The choice of this particular observation was driven also by the
existence of a multiwavelength campaign (Mirabel {\em et al.}, 1998), that
could allow a richer discussion between QPOs, disk instabilities and
ejections of plasma, since several authors have now linked the changes
of luminosity to disk instabilities ({\em e.g.} Feroci \etal, 1999; 
Fender \etal, 2001, and references therein). 
We have thus analyzed two such cycles, essentially repeating the
analysis of SM97 and turning our attention to the frequency-radius
correlation.

We will present the observations of \J1655 and \G1915 in sections
\ref{subsec:J1655} and \ref{subsec:G1915} respectively.  We will then in
section \ref{subsec:res1915} discuss and compare the results, which 
will be
then be confronted with a theoretical model in Paper II. We will also
discuss the different behaviors observed when the disk color radius
takes anomalously small values, leading to a discussion of disk
properties and to a possible explanation for these very small values.
\section{Data reduction and analysis}
\label{sec:data}
\subsection{\J1655}
\label{subsec:J1655}
\begin{figure}[htbp]
\includegraphics[width=\columnwidth]{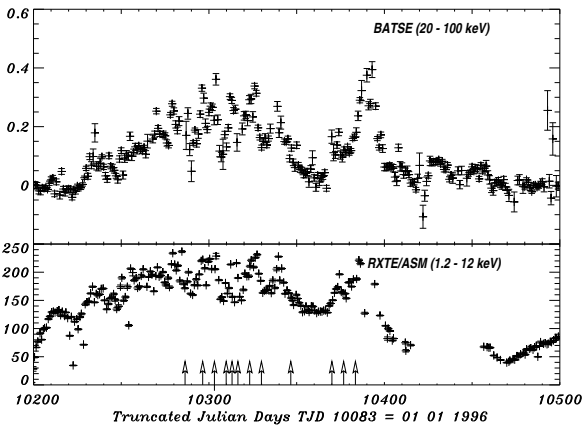} 
\caption{GRO J1655-40 light curves during 1996, obtained at high
energies with BATSE
($20-100$ keV) and at lower ones with the RXTE All Sky Monitor ($1.2-12$
keV).  Arrows show the dates where low-frequency QPO were observed.}
\label{fig:lightcurve}
\end{figure}

GRO J1655-40 (Nova Scorpii 1994) was discovered in 1994 by the Burst and
Transient Source Experiment (BATSE) (Wilson et al.  1994).  Its optical
counterpart was discovered in 1995 (Bailyn et al.,1995), and the orbital
parameters gave a primary mass of $7.02 \pm 0.22 M_\odot$ (Orosz \&
Bailyn, 1997; van der Hooft et al., 1998), placing it well above the $3
M_\odot$ maximum mass for neutron stars.  Its distance is estimated at
$3.2 \pm 0.2$ kpc (Hjellming \& Rupen, 1995), with an inclination angle
$\sim 69.5^{\circ}$.  GRO J1655-40 is also known as one of the few
galactic microquasars, since radio observations show two extended radio
lobes with superluminal motions (Hjellming \& Rupen, 1995).
In order to be able to compare directly the results on the disk color
radius with the ones presented below for \G1915, we have re-done the
data reduction and analysis of the series of twelve public observations
(AO1) of \J1655 during its 1996 outburst (see figure
\ref{fig:lightcurve}), already presented by SMR, when QPO were 
observed. 
Figure \ref{fig:lightcurve} shows the BATSE and RXTE ASM light curves
during that period, and the dates where QPO were observed.  Table
\ref{table:J1655} gives the corresponding disk parameters.

For our analysis the RXTE spectra were obtained using only the
Proportional Counter Array (hereafter PCA).  We used the standard 2 data
only when the offset pointing was less than $0.02^{\circ }$ and the
elevation angle above $10^{\circ }$; in almost all cases all the PCU
were ``on'' during the observation, but we chose to use only the
spectra extracted from PCU 0 and 1 separately, and then fit them
together in XSPEC. Given the high flux of the source this allows us to
get sufficient statistics while limiting the effects associated with
the various PCUs.
\begin{figure*}[htbp]
\includegraphics[width=\columnwidth]{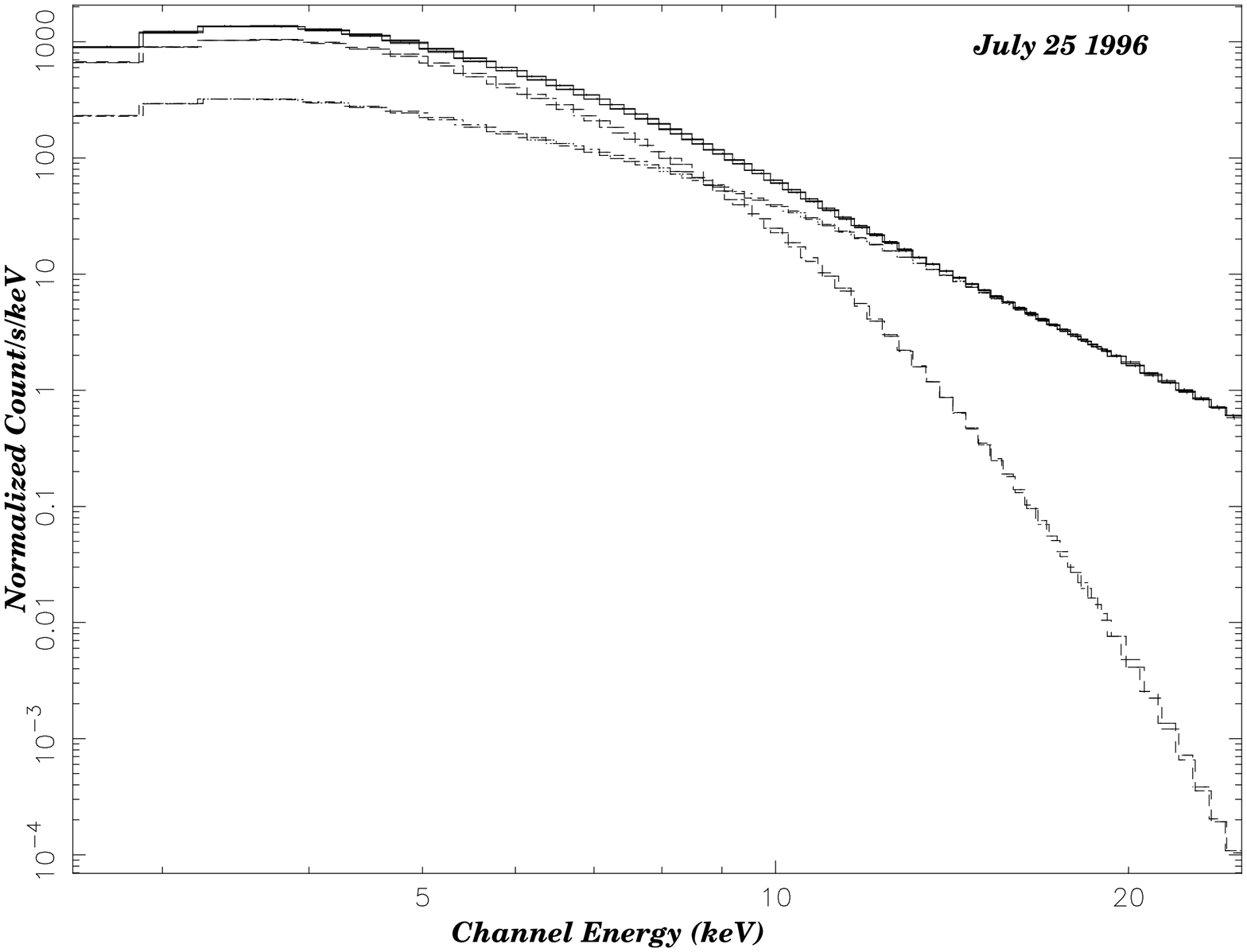} 
\includegraphics[width=\columnwidth]{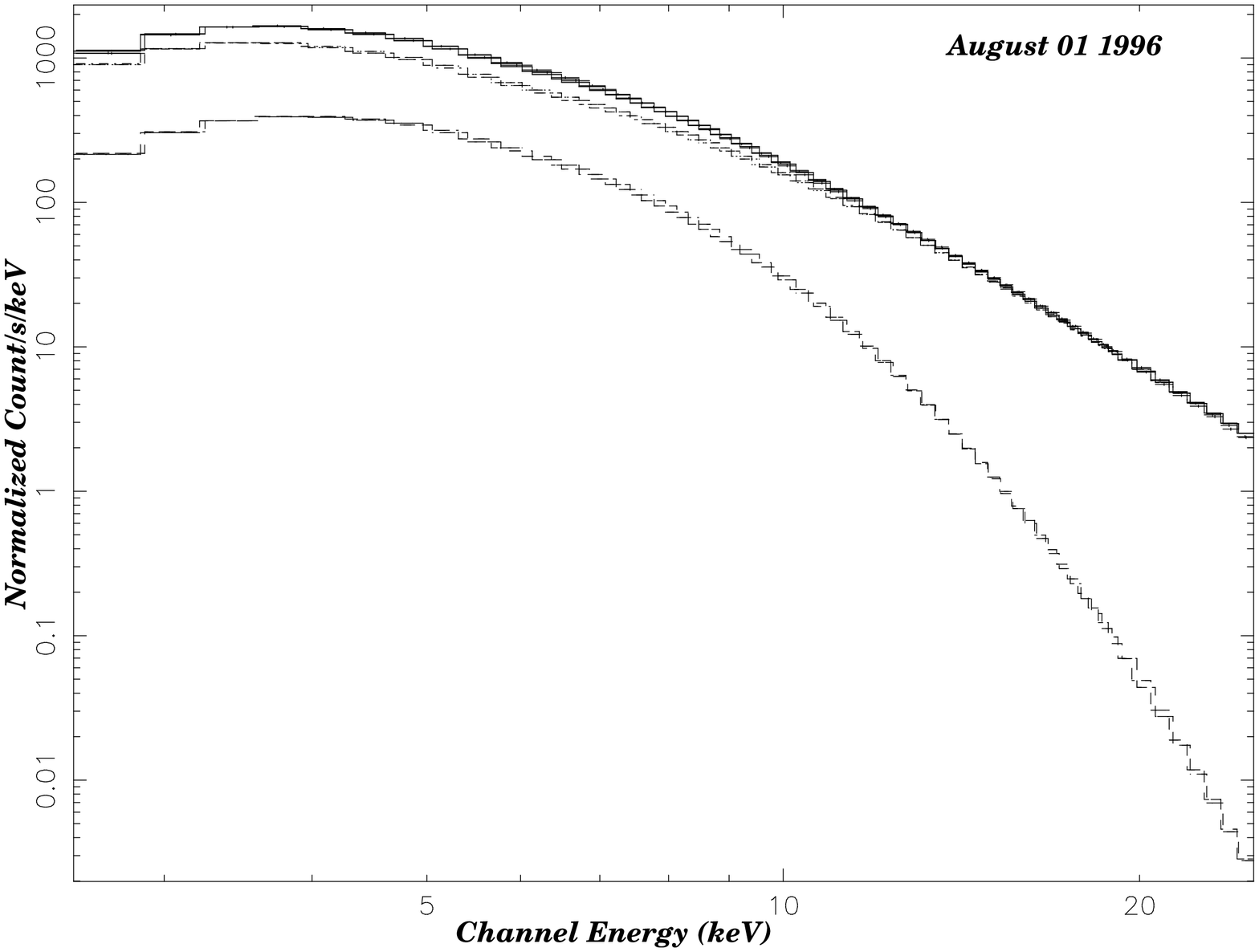} \\
\caption{Spectral fits for the first two observations of \J1655. On 
July 27 the black-body part dominates at low energy, allowing a 
satisfying extraction of disk parameters. On August 1 the coronal part 
dominates at all energies.
The plots show (top curve) a superposition of spectra from PCU's 0 \& 
1, 
and (lower curves) the spectral components (multicolor black body and 
power-law tail) obtained from the fit.  The X axis is in keV, and the Y
axis is in Counts/sec/keV.}
\label{fig:spectra1655}
\end{figure*}
The data reduction/extraction was made following the standard RXTE data
reduction. In order to compare precisely with the results of SMR, we 
have used FTOOLS package 4.1, but checked that package 5.0 gave nearly 
identical results.

We first generated spectra using all the PCU layers, which gives a
higher count rate, and then extracted spectra only from the top layer;
the fits parameters deduced with the two methods presented no
significant differences, and since there was, for this source, no
particular need to get the highest count rate (the source was
sufficiently bright, and the exposure sufficiently long \correc{that the data
 had small statistical errors)}, the results presented here are obtained
with the later method.The background was estimated using the PCABACKEST
tool version 2.1, and we chose to ignore channel energies bellow $2.5\
keV$, and above $20\ keV$, in order to use the best detector energy band
(choosing this band, we avoided the uncertainties due to the response
matrix at lower and higher energies, and the problems due to background
subtraction near $25\ keV$ -- see also SMR); we then added a systematic
error of $1.5\%$ on each data point.  The response matrices were
generated using the PCARSP tool version 2.38, and we deadtime correct
the spectra following the RXTE cook book method.  The spectral fits were
obtained using XSPEC version 10., using the multicolor disk blackbody
model with a power law tail at higher energy, and a hydrogen column
absorption fixed at $N_{H}=0.9 \times 10^{22} cm^{-2}$ as found with
ASCA (Zhang et al., 1997a).  We have found that no other contributions 
to
the spectrum (e.g iron line) needed to be introduced in the fit, giving
results consistent with those of SMR while reducing the number of fit
parameters.  On the other hand, since there is little uncertainty on the
QPO centroid frequencies, we have not re-done the power spectra
analysis, and we use here the frequency values given by SMR.

Table \ref{table:J1655} shows the results of our fits.  Column 7 shows
the ratio ({Disk black body flux}/{total flux}) which, according to MFR,
we will first use to discriminate between two different spectral states
of the source.  For our purpose, the total flux is obtained by
extrapolating the spectra to 50 keV. Table \ref{table:J1655} shows that
the observations clearly come in two categories: one where the ratio
{Disk black body flux}/{total flux} is lower than .5, so that according
to the criterion of MFR the disk radius value should be considered as
unreliable, and where indeed the radius is {\em always}
unrealistically low; and
one where the ratio is higher than .5 and where the obtained radius
appears to be more trustworthy: it lies in the range we expect for a
disk around a black hole, and gives more consistent values.

Figure \ref{fig:spectra1655} shows the spectra for the first two
observations, typical of these two behaviors: on July 25 the disk
emission dominates at low energy, allowing a satisfying spectral fit. 
On August 1 the coronal emission dominates at all energies, and results
in an unreliable fit.  Figure \ref{fig:correl1655} is the resulting
plot, for the data points we retain, of the QPO centroid frequency
versus the disk color radius, where $r_{col}$ is given in \correc{km,
with $R^{\star}=((r_{col} /km)/(D/10kpc))^{2}\cos\theta$ $R^{\star}$ being
the XSPEC dimensionless parameter}.  
Comparing with figure 2 of SMR, we note that the
points we exclude correspond to higher temperatures and lower disk 
flux. 
It is quite remarkable that they are also the observations where a
higher frequency QPO is simultaneously observed.  This might confirm
that indeed the disk is in a different state in the observations we
retain and the ones we reject, but we will not attempt to explore an
explanation, which would clearly require reaching a much better
understanding of the disk behavior than our present one -- although in
our discussion we will contrast this with the behavior of \G1915.  On
the other hand, with the data points we retain, we confirm the general
tendency noted by SMR that, contrary to other sources, the QPO frequency
increases with increasing disk radius in \J1655.

\begin{table*}[htbp]
\begin{tabular}{|c||c|c|c|c|c|c|c|}
\hline
$\#$ & Date & $T_{col}$ (keV) & $R_{col}$ & $\alpha$ & Powerlaw & Ratio& 
$\chi_{\nu}^{2}$ (91 d.o.f)\\
\hline
\hline
$1$ & $07$-$25$-$96$ & $1.272 \pm 3.6\times 10^{-3}$ & $20.62 \pm 
0.15$ & $2.589 \pm 0.022$ & $15.93 \pm 0.98$ & $0.64$ & $0.69$\\
$2$ & $08$-$01$-$96$ & $1.63 \pm 1.4\times 10^{-2}$ & $8.41 \pm 0.28$ 
& $2.49 \pm 0.012$ & $50.68 \pm 1.78$ & $0.21$ & $0.76$\\
$3$ & $08$-$06$-$96$ & $1.53 \pm 8.7\times 10^{-3}$ &$ 10.90 \pm 
0.23$ & $2.65 \pm 0.013$ & $47.93 \pm 1.856$ & $0.31$ & $0.7$\\
$4$ & $08$-$15$-$96$ & $1.247 \pm 3.7\times 10^{-3}$ &$ 20.69 \pm 
0.15$ & $2.56 \pm 0.024$ & $13.58 \pm 0.9$ & $0.65$ & $0.98$\\
$5$ & $08$-$16$-$96$ & $1.27 \pm 6.6\times 10^{-3}$ & $16.88 \pm 
0.27$ & $2.59 \pm 0.016$ & $34.40 \pm 1.578$ & $0.36$ & $0.82$\\
$6$ & $08$-$22$-$96$ & $1.25 \pm 5\times 10^{-3}$ & $19.10 \pm 
0.20$ & $2.6 \pm 0.017$ & $26.79 \pm 1.26$ & $0.47$ & $0.62$\\
$7$ & $08$-$29$-$96$ & $1.59 \pm 1.2\times 10^{-2}$ & $9.41 \pm 
0.27$ & $2.54 \pm 0.013$ & $50.28 \pm 1.86$ & $0.24$ & $0.68$\\
$8$ & $09$-$04$-$96$ & $1.25 \pm 4\times 10^{-3}$ & $20.56 \pm 
0.17$ & $2.61 \pm 0.02$ & $19.50 \pm 1.0$ & $0.59$ & $0.72$\\
$9$ & $09$-$20$-$96$ & $1.23 \pm 3.6\times 10^{-3}$ & $21.10 \pm 0.14$ 
& $2.49 \pm 0.023$ & $9.94 \pm 0.65$ & $0.69$ & $0.7$\\
$10$ & $10$-$15$-$96$ & $1.29 \pm 3.5 \times 10^{-3}$ & $20.80 \pm 0.13$ 
& $2.36 \pm 0.028$ & $7.098 \pm0.56$ & $0.74$ & $0.6$\\
$11$ & $10$-$22$-$96$ & $1.28 \pm3.5\times 10^{-3}$ & $21.13 \pm 
0.13$ & $2.48 \pm 0.025$ & $10.79 \pm 0.76$ & $0.70$ & $0.62$\\
$12$ & $10$-$27$-$96$ & $1.32\pm5\times10^{-3}$ &$17.84 \pm 
0.21$ & $2.64\pm 0.017$ & $34.31\pm1.66$ & $0.46$ & $0.7$\\
\hline
\end{tabular}
\caption{Best fit parameters for the observations of \J1655. 
$R^{\star}=((R_{col} /km)/(D/10kpc))^{2}\cos\theta$, where $\theta$ is
the inclination angle.  Powerlaw normalization is in unit of photon/
$cm^{2}$ /s at 1 keV.}
\label{table:J1655}
\end{table*}

\begin{figure}[htbp]
\caption{Plot of $\nu{QPO}$ vs $r_{col}$ for GRO J1655-40. X axis is in 
km, Y axis is in Hz.}
\label{fig:correl1655}
\end{figure}

\subsection{\G1915}
\label{subsec:G1915}
We thus turn to the microquasar \G1915.  This source has been the
object of very frequent observations, many of them showing the
low-frequency QPO and strong variability.  In particular, the very
exhaustive analysis of Muno {\em et al.} (1999) shows the QPO, in
different spectral states, and over a large range of
frequencies.  Furthermore, SM97 had shown that, during the $\sim$ 30
minutes cycles of \G1915, the QPO could be tracked at a varying
frequency, together with a color radius varying by a factor $\sim$ 5,
during the ``low and hard'' part of the cycle.  These cycles have been
extensively studied; multiwavelength observations (Mirabel {\em et al.},
1998; see also Chaty, 1998) have shown them to be associated with
supra-luminal ejections, and they turn out to be the most frequent type
of behavior of that source after the steady ones (Muno {\em et al.},
1999; Belloni {\em et al.}, 2000).  Observing the correlated changes of
radius and QPO frequency during such events may limit the influence of
other effects, such as the mass accretion rate etc., and provide a way
to analyze the disk history during such events.  This could then
complement the analysis of Belloni {\em et al.} (2000), based only on
color-color diagrams.  

\begin{figure}[htbp]
\includegraphics[width=\columnwidth]{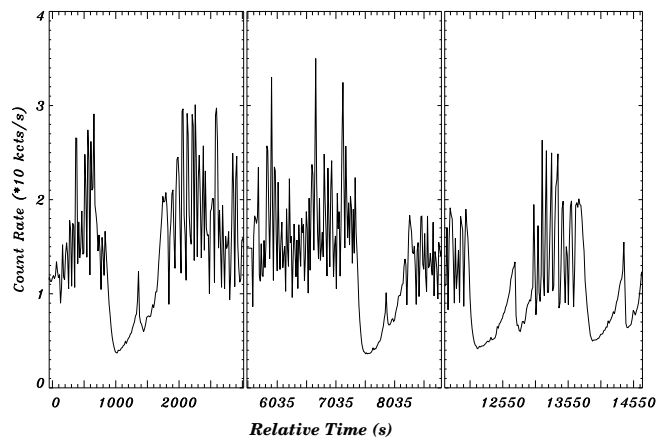} 
\caption{Lightcurve of \G1915 during the whole observation of 1997 Sept. 
9.  X axis is in seconds since the beginning of the observation, Y range
is in count/s.}
\label{fig:GRS1915light}
\end{figure}

Figure \ref{fig:GRS1915light} shows three consecutive cycles observed by
RXTE on 1997 Sept.  9.  The first cycle is the one shown in detail by
SM97, while the second cycle was the object of the multiwavelength
observations of Mirabel {\em et al.} (1998).  We have thus chosen to
analyze these two consecutive cycles: we repeat the analysis of SM97 for
the first cycle, using for the spectral fits and color radius derivation
the same procedure as for \J1655.  In order to confirm the results we
obtain, we repeat the same analysis for the following cycle.  We have
also made a rapid analysis showing that the same behavior is observed
during the third cycle.  We treat only the time interval over which the
QPO was observed by SM97, {\em i.e.} starting near the transition from
the high/soft to the low/hard state, and ending at the intermediate peak
halfway through the low state.  At that time the QPO disappears, and the
power-law emission decreases dramatically (Mirabel {\em et al.}, 1998;
Chaty, 1998).
\subsubsection{Spectral analysis}
Our goal was to track the variations of the QPO frequency and of the
disk color radius, while they vary rather rapidly.  We thus reduced the
time interval over which the spectral fit and the QPO frequency
determination were obtained, to the minimal value compatible with
sufficient statistics.  We used the standard 2 data accumulated over 16s
intervals; although the source
occasionally varies on shorter time scales, we have checked that this
does not affect the results we present.  Due to the poorness of the
statistics, we extracted spectra from all available PCU (though only
PCU 0 to 3 were ``on'' during the whole observation), and all layers
to get the most possible incoming flux.  The spectra, after dead time
correction and background correction, were first fitted, as for GRO
J1655-40, between $2.5$ and $20$ keV in XSPEC with the standard model
consisting of a multicolor disk blackbody and a power law tail at higher
energy, taking into account a low energy absorption with $N_H$ fixed at
$5.7\times 10^{22} cm^{-2}$ ( Markwardt {\em et al.}, 1999).  The fits
were not satisfactory, with an average reduced $\chi$ square $\sim 
2.5$. 
An analysis of the residual showed a large difference between the
spectra and the model used around 6 keV, so that in a second pass we
added to the model an iron emission line, with centroid energy found at
$\sim 5.9$ keV. \\

As for \J1655, the coronal emission dominates the light curve in some
observations.  We thus choose the following procedure: we distinguish in
the observations a high state and a low state; the latter is defined by
the combination of a low temperature ($\simlt 1$ keV) and a large
radius.  In this state, the coronal contribution dominates but the
correction of MFR (which was derived for high states) should not apply,
and the color radius is much larger than at other times, so that we
retain these points - which correspond to most of the intervals where
the QPO was present.  In the high state on the other hand, the radius is
much lower and we apply the same criterion as above, retaining only data
points where the black-body contribution exceeds half of the total
(extrapolated to 50 keV), although we find that in general the radius
values for the rejected points are more consistent than in \J1655.  The
resulting parameters and classification are presented in Table
\ref{table:1915spec}.
\subsubsection{Temporal analysis}
\begin{figure*}[htbp]
\centering
\includegraphics[width=\columnwidth]{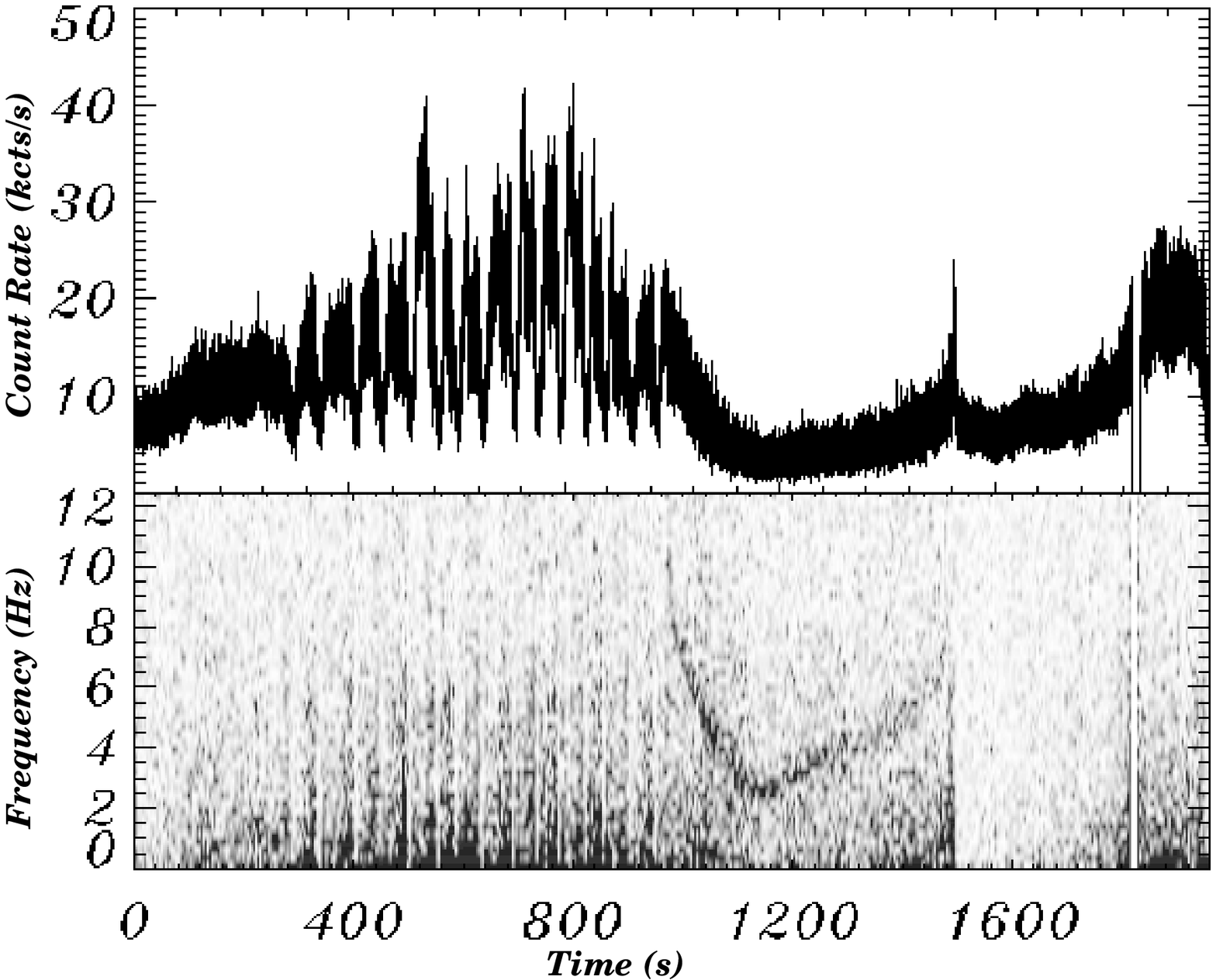} 
\includegraphics[width=\columnwidth]{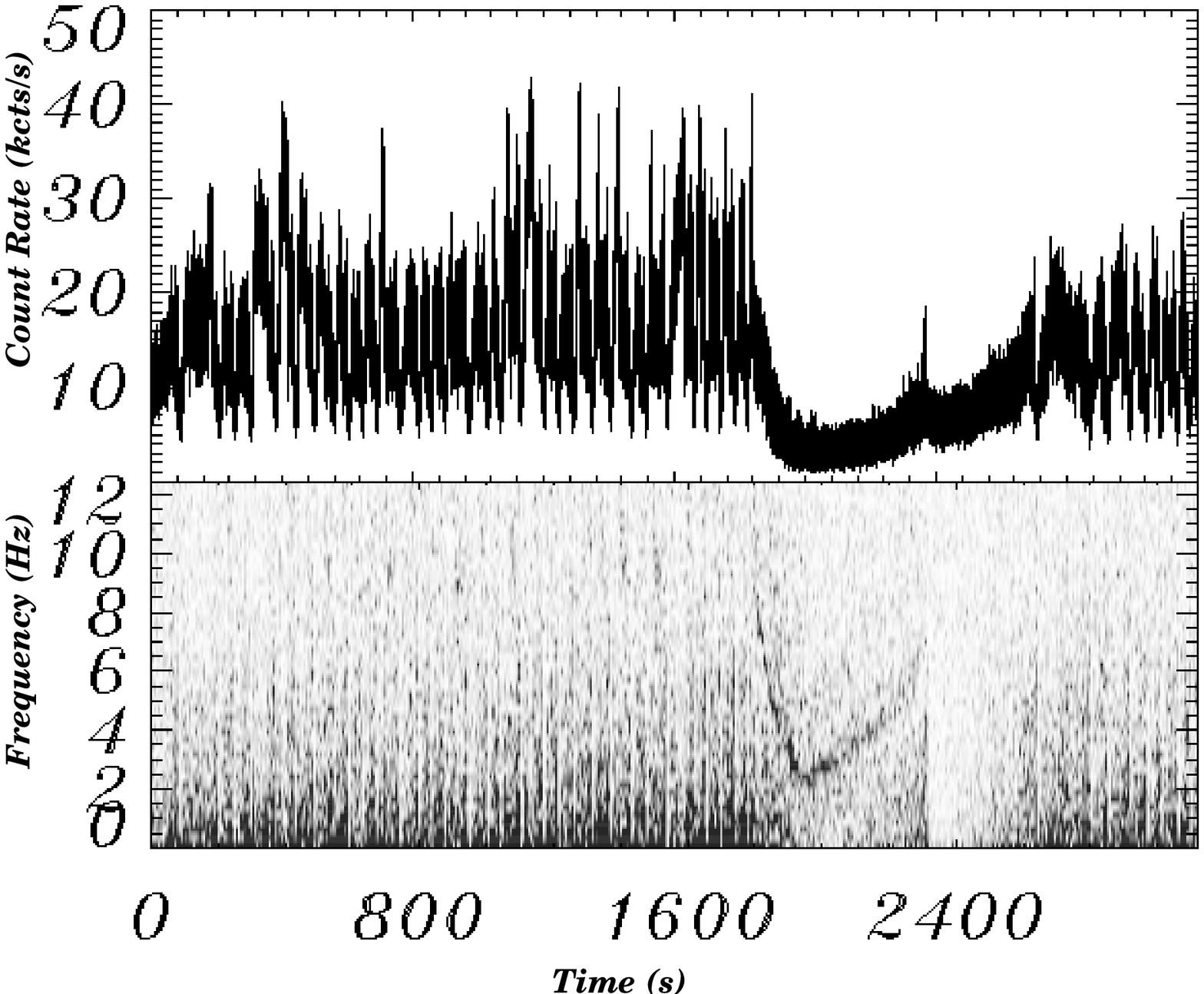} 
\caption{\itshape{Dynamical power spectra of GRS 1915+105.  The left
and right panels correspond respectively to the first and second cycles
analyzed.  The X axis is is seconds, with the zero corresponding in each
case to the beginning of accumulation.}}
\label{fig:powspectra}
\end{figure*}

As we wanted to follow the QPO in the range 2-12 Hz found in SM97, 
we generate power spectra for the first cycle of figure
\ref{fig:GRS1915light} using the binned data, with 8 ms resolution, and
the event data with time resolution of 62 $\mu$s collected during 4s
ccumulation, after background correction.  Actually the background is
so low that retaining it would change the results only by less than
$\sim\ .5\%$.  From this we generated a dynamical power spectrum of the
source, repeating the analysis of SM97.  Our result, shown on figure
\ref{fig:powspectra} (left), confirms theirs: the shaded, U-shaped
pattern seen on the lower part of figure \ref{fig:powspectra} between
times $\sim 1000$ and $\sim 1500$, shows the QPO frequency and its
evolution during the cycle.  We then repeated this analysis for the
second cycle, confirming the generality of this behavior (figure
\ref{fig:powspectra}, right).

\begin{figure*}[htbp]
\includegraphics[width=\textwidth]{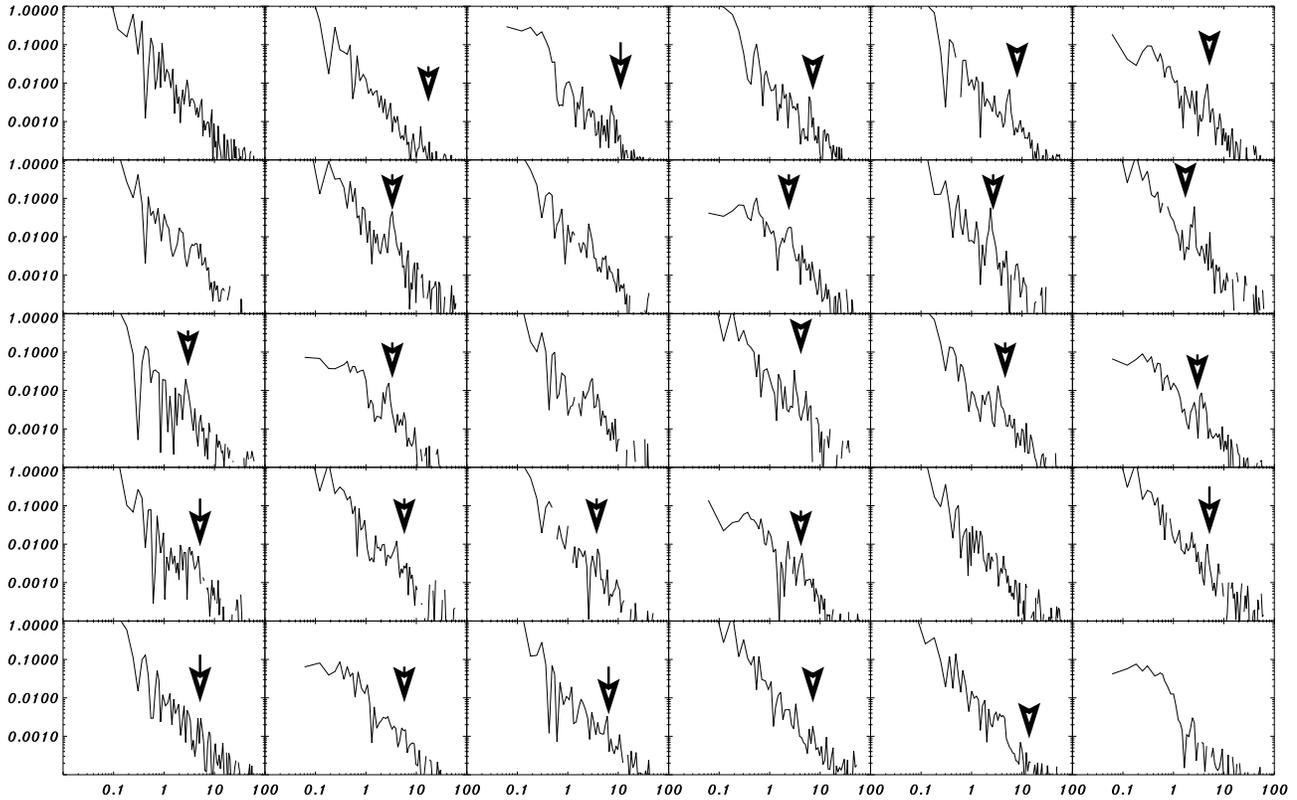} 
\caption{Power
Spectra from the second cycle; top left corresponds to the first time
interval analyzed in this cycle, and bottom right to the last one. 
Arrows indicate the position of the QPO found from the fit in each
interval, using the dynamical power spectrum (figure
\ref{fig:powspectra}) as a guide to reduce ambiguities.  Even with this
help, the limited statistics makes it difficult or impossible in some
cases to correctly identify the QPO.}
\label{fig:allpow1915}
\end{figure*}
\begin{figure*}[htbp]
\includegraphics[width=\columnwidth]{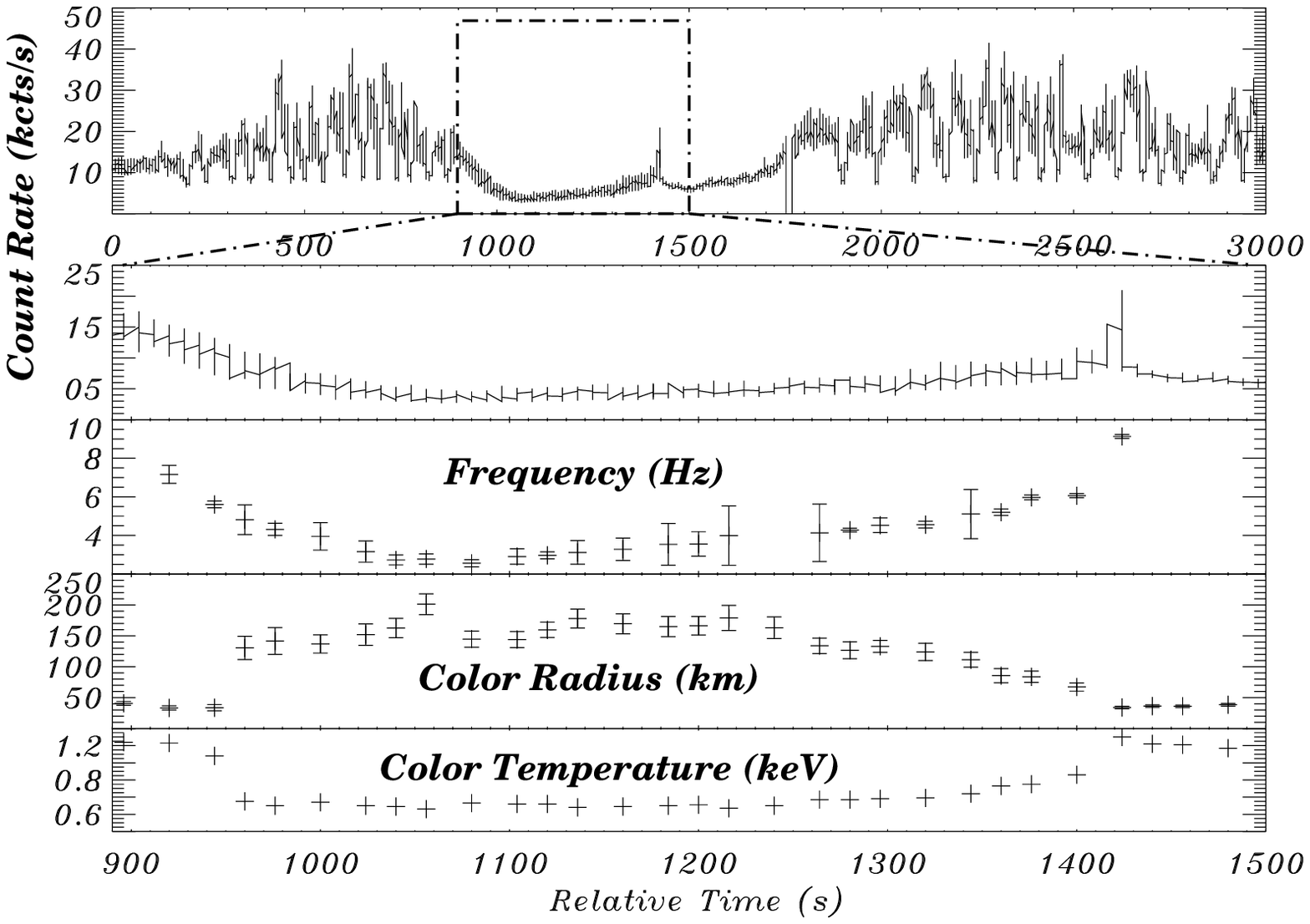} 
\includegraphics[width=\columnwidth]{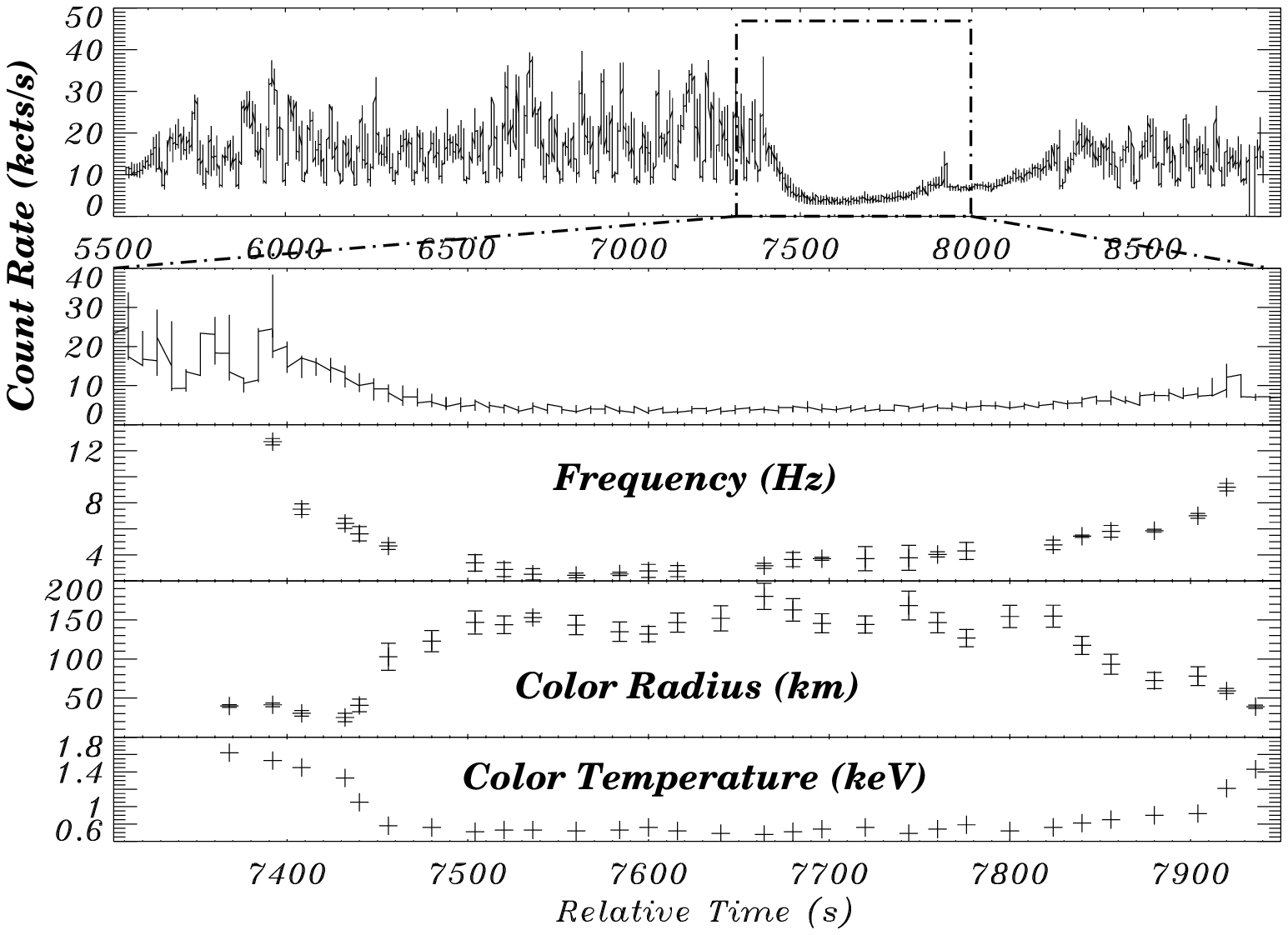} 
\caption{Plot of the timing evolution of the source flux during the
first (left), and second (right) cycle.  Subpanel is a temporal zoom on
the interesting timing region, with, under the zoomed lightcurve, the fit
values of the QPO frequency and the disk radius.  Error bars are one
$\sigma$ confidence level.}
\label{fig:superpos}
\end{figure*}

The second step was then to track as precisely as possible the QPO
frequency, and to correlate it with the disk color radius extracted with
the same process used for \J1655 in section \ref{subsec:J1655}.  We do
this using the same time resolution data, but with exposures of 16s. 
Power spectra were then generated using POWSPEC version 1, after
background subtraction and collimator correction; they were then fitted
with a model consisting of a power law with average index $\sim -1.5$,
plus a lorentzian whose line center is taken as the QPO centroid
frequency; in some cases several lorentzians were required to fit 
the 16s power spectrum; 
we then chose the one closest to the frequency at which power was strong 
in power spectra of neighboring intervals.  The trend of the usually strongest 
feature is easily picked out from the dynamical power spectrum.
The choice of this 16 second interval
for the temporal analysis is a compromise between conflicting
constraints: looking at variations of the disk radius and the QPO
centroid frequency on short enough time scales (without averaging on too
long time intervals over which the source varies rapidly), having enough
flux to get sufficient statistics on each data point, and getting the
QPO frequency with good accuracy.

\subsection{Results}
\label{subsec:res1915}
 \correc{The individual power spectra, obtained from interval \# 2, are shown 
in Fig. \ref{fig:allpow1915}.}
  One sees that, with the time intervals of 16s., the QPO
can most often be identified and properly found with our fitting
procedure.  
Figure \ref{fig:superpos} shows superpositions of our results for the disk 
color radius, the QPO frequency, and the disk temperature during both cycles.
 Figure \ref{fig:1915plot} shows, for both cycles, the correlation
between the QPO frequency and the disk color radius.  At large $r_{col}$
the QPO frequency decreases with increasing $r_{col}$, as discussed by
Sobczak {\em et al.} (2000) for \X1550; this is also the general
tendency for this correlation, noted by SM97 during one cycle, and more
generally by Muno {\em et al.} for an ensemble of observations (in
particular for similar 30 min.  cycles).\\

 It would be tempting to interpret the points at lower radius as a turnover, 
indicating a reversed correlation as seen in \J1655.  
However, returning to Fig.
\ref{fig:superpos}, one sees that this interpretation must be excluded:
the points at lowest radius correspond to the very first manifestations
of the QPO. At this time the count rate has not yet started to 
decrease (this might indicate that the appearance of the QPO is not a 
consequence of the transition, but possibly its cause),
while the QPO frequency decreases monotonously; the color radius first
decreases before it starts growing, when the count rate decreases,
marking the transition to the low state: this causes the reverse
correlation seen in Fig. \ref{fig:1915plot}.  On the other hand,
returning to the data points {\em in the high state, before the onset of
the QPO} one sees that in its variations the color radius consistently
shows a fixed lower bound, which probably marks the Last Stable Orbit.\\
The data points showing an inverse QPO frequency-color radius
correlation lie at lower color radius, and at these times the disk
temperature decreases, as predicted by MFR. It is thus very likely that,
in the spectral fits for these points, the correlation between the color
radius and the disk inner radius has become anomalous.
\section{Discussion}
The possibility, which will be explored in Paper II, of a theoretical
explanation has led us to reconsider the contrasting correlations,
observed by SMR in \J1655 and \X1550, between the disk color radius and
the QPO frequency.  First guided by the arguments of MFR, we have
rejected data points for \J1655 where the measure of the disk color
radius was dubious.  The major result of this work is thus that,
though we have used the most stringent criterion to select ``good''
observational data points, the reverse correlation remains: the QPO
frequency decreases with decreasing radius in the case of \J1655.\\
 In the case of \X1550 the data, for the observations where a QPO is
seen, always correspond to a low value of the ratio disk black
body/total flux, and may not be reliable according to the criterion of
MFR. We have thus turned to \G1915, because previous work by SM97 and
Muno {\em et al.} (1999) had shown that, during the frequently observed
30 min.  cycles of that source, the QPO frequency and disk color radius
varied in a correlated manner over a large interval.  We have done a
complete analysis of two of these cycles, sampling data over the minimal
time interval compatible with proper statistics in order to follow as
precisely as possible the rapid variations of the source.  This does not
show a reverse correlation as in \J1655.  Data points obtained
at the onset of the QPO would seem to indicate it, but a detailed
analysis shows that they are probably incorrect (the spectral fit
returning an anomalous value of the disk color radius), so that we have
to rule them out.

\begin{figure}[htbp]
\caption{Plot of $\nu_{QPO}$ vs $r_{col}$ for GRS1915+105 for the 
first (squares) and second (circles) cycles; error bars shown on figure 
\ref{fig:superpos} are not displayed here, for the sake of clarity.}
\label{fig:1915plot}
\end{figure}

However the measure of the radius is not an absolute one.  It is
obtained from a multicolor black-body + power law tail model (plus an
iron line when needed), involving a number of hypotheses (such as
axisymmetry of the disk emission, and $\sim r^{-3/4}$ radial temperature
profiles), and should be submitted to a number of corrections (such as
electron diffusion or spectral hardening factor) which we have not tried
to include here.  Furthermore, MFR have recently shown that effects due
to the finite disk thickness could result in a systematic underestimation
of the radius, when it becomes small.  It is remarkable here that, for
the \G1915 data we have rejected, the spectral fit returns a color
temperature which is {\em lower} than usual, and in particular lower
than at times just preceding the onset of the QPO, although the radius
decreases: this is compatible with the predictions of MFR, whose 
mechanism relies in part on the dissipation of part of the accretion 
energy in the corona rather than the disk (so that the disk cools down).

On the other hand, the data we have rejected for \J1655 show a very
different behavior, also frequently observed in the low and steady state
of \G1915 (Muno {\em et al.}, 1999): there, the color radius takes
extremely small values (sometimes very few kilometers) while the color
temperature becomes high.  This is not the behavior predicted by MFR,
and it is quite unlikely that corrections of this type could explain so
strongly anomalous values of the radius: the lowest radius value found
by Sobczak {\em et al.} (2000) is about 4 times lower than the ones we 
retain.\\

However, we note that, if the emissivity is close to unity, the spectral
fits actually measure (essentially by the intensity of the black body
emission) the {\em size} of the emitting region: a simplistic
view of the multicolor blackbody spectrum used in the spectral fits is
that the peak of the emission corresponds to the highest temperature in
the disk, and the total intensity to the area of the emitting region at
that temperature, all this weighted by the temperature distribution in
the disk.  In \J1655 we find that the color radius can strongly vary,
while the total luminosity remains approximately constant.  The most
obvious explanation would then be that the total accretion energy
dissipated in the disk does not vary much, but that the area of the
region where it is dissipated does.  This leads us to suggest that the
anomalously low measured radius, {\em together with a high temperature},
could be interpreted by the presence in the disk of a hot point or a
spiral shock, where a substantial fraction of the accretion energy would
be dissipated.  A spiral shock could result from the non-linear
evolution of the AEI, just as the gas forms shocks in the arms of spiral
galaxies.\\
The discussion about the exact ratio $r_{col}/r_{in}$, related to
the hardening factor by $r_{in}=r_{col} \times f^2$, with
$f=T_{col}/T_{eff}$ , being the hardening factor (Ebisawa {\em et al.},
1994) is a very complex one.  It appears to be dependent on many
physical parameters, as shown {\em e.g.} by Shimura \& Takahara (1995),
for a Schwarschild Black Hole.  Although both \G1915 and \J1655 are
thought to be a Kerr Black Hole, we adopted the best result found in their
work, $f=1.9$, and tentatively estimated the position of the last stable
orbit, using $R_{LSO}\sim \frac{3}{5} \times 0.8 R_{col}f^2$ (Ebisawa
{\em et al.}, 1994).  We chose for each source the lowest ``good'' value of
$R_{col}$ as the one approaching $R_{LSO}$ the most. In the
case of \J1655, this leads to an estimation of $R_{LSO}\sim36$ km,
inconsistent with a $\sim 7M_\odot$ Schwarschild Black Hole, where
$R_{LSO}$ would be of the order of $63$ km.  This large difference can
be due to two effects.  First, the value we adopt is not the exact
measure of the last marginally stable orbit, but is really a rough
estimation of it (we at least expect it to be closer to the compact
object).  The second one, which may be the major effect, is that this
source, in particular, is expected to be a maximally rotating Black Hole
(Strohmayer, 2001), where  $R_{LSO}=17$ km, which is, then, much more
consistent (Paper 2).  The case of \G1915 can now be explored, since
Greiner (2001) has estimated the mass of the Black Hole to be close to
$14M_{\odot}$.  The result found ($R_{LSO}\sim59$ km) , as for \J1655,
is inconsistent with a Schwarschild Black Hole of $14M_{\odot}$, but
would perfectly fit the case of a maximally rotating Kerr Black Hole of
such mass (see Paper 2). \correc{We also note as an interesting possibility to discuss 
the mass of the Black Hole the work of di Matteo and Psaltis (1999), 
which should be enriched by recent measurements of this mass.} \\
But we have to exercise here, extreme caution, since :
\begin{itemize}
\item all the calculations in Shimura \& Takahara (1995), in particular
the determination of the best value of the hardening factor, are made 
for a Schwarschild black hole; 
\item all the calculations, presented in Ebisawa {\em et al.} (1994), and
Shimura \& Takahara (1995), are done for a Shakura \& Sunyaev (1973) 
$\alpha$-disk, where energy and angular momentum are locally
dissipated in the disk under the effect of viscous stresses; whereas in the 
framework of the AEI, only a small part of the energy warms the disk.
Most of it is transported toward the corotation radius (Tagger \& Pellat, 
1999; Paper 2, Varni\`ere {\em et al.}, {\it{in Prep.}});  
 \item the Kerr metric may also affect the relation of Ebisawa 
\etal (1994), between $R_{LSO}$, the radius of the Last Stable Orbit, and 
the measured $R_{col}$;
\item although in the case of \J1655 the black hole is thought to 
be rapidly rotating, this is still unsolved for \G1915; 
its temporal behavior does not allow a firm conclusion (Strohmayer, 2001) 
(one has to note, nevertheless, that previous studies give preference 
to a rapidly rotating black hole (Zhang \etal, 1997).
\end{itemize}

The relevance of these calculations is thus a matter of debate,
and  we cannot draw conclusion on the value $f=1.9$ for the hardening factor, 
or on the rotation of the black holes. 
It is however remarkable that 
the independent method presented in Paper 2 gives a good range of masses, 
for both sources, in the cases where both are spinning rapidly.\\
Our conclusions must be considered as tentative, given the uncertainties
on the measure of the disk color radius and its relation with its inner
radius, and given the other effects associated with the disk parameters,
such as the amplitude and the radial distribution of the magnetic field. 
On the other hand, a coupled investigation of the low frequency QPO and
of the disk spectral properties seems to be a most promising way to
probe the physics of the inner accretion disk in black hole binaries.
\begin{acknowledgements}
The authors wish to thank many experts of the field for helpful
discussions and comments.  In particular we thank C. Markwardt
(including help with his IDL library site and advice on producing
dynamical power spectra), S. Chaty, F. Mirabel, E. Morgan, M. Muno, and
R. Remillard.  We also thank all the people contributing to the GSFC
public database existence and update, and the helpful and constructive
comments of both referees.
\end{acknowledgements}
\bibliographystyle{plain}

\begin{table*}[tbp]
\begin{tabular}{|c|c|c|c|c|c|c|c|c|c|c|}
\multicolumn{9}{c}
{\itshape Table 2} \\
\hline
interval & $T_{in}$  & $r_{col}$ & $\alpha$ &Powerlaw& $\chi_{\nu}^2  $
& QPO freq. &$ Q$ & $\chi^2$ (d.o.f)&$\frac{F_{dbb}}{F_{tot}}$& 
Spectral state  \\
 (s)  &  (keV) &  (km)  &   &  & (40 dof)  & (Hz) &   & & &
 \\
\hline
\hline
116403340-116403356 & 1.44 & 40.76 & 2.74 & 47.84 & 0.63 & 
**** &   & &{{{0.44}}} & {{{high state}}}\\
116403360-116403376 & 1.43 & 33.26 & 2.79 & 50.3 & 
0.78  & 7.16 & 15.42 & 23.51 (41) & {{{0.30}}}& {{ {high 
state}}}\\
116403380-116403396 & 1.28 & 33.66 & 2.79 & 39.03 & 
0.85 & 5.60 & 32.27 & 31.70 (35) & {{{0.2}}} & {{{high state}}}\\
116403400-116403416 & 0.75 & 130.55 & 2.63 & 24.45 & 1.29 & 
4.81 & 6.25 & 35.5 (51) & {{{0.29}}} &{{{low state}}}\\
116403420-116403436 & 0.70 & 141.56 & 2.68 & 22.15 & 1.11 & 
4.31 & 13.77 & 16.90 (35)& {{{0.29}}} &{{{low state}}}\\
116403440-116403456 & 0.74 & 136.82 & 2.52 & 12.7  & 0.95 & 
3.95 & 5.56 & 24.98 (31) & {{{0.41}}}&{{{low state}}}\\
116403460-116403476 & 0.70 & 151.96 & 2.57 & 10.61 & 1.01 & 
3.16 & 5.74 & 20.02 (31)&{{{0.48}}}&{{{low state}}}\\
116403480-116403496 & 0.69 & 162.55 & 2.42 & 6.67  & 0.7 & 
2.73 & 10.11 & 18.29 (27) & {{{0.59}}}& {{{low state}}}\\
116403500-116403516 & 0.66 & 201.13 & 2.34 & 5.21  & 0.68 & 
2.77 & 10.28 & 19.42 (35) & {{{0.68}}} & {{{low state}}}\\
116403520-116403536 & 0.73 & 144.69 & 2.43 & 6.21  & 0.83 & 
2.56 & 13.83 & 19.95 (35) & {{{0.63}}} & {{{low state}}}\\
116403540-116403556 & 0.72 & 143.84 & 2.53 & 8.78  & 0.97 & 
2.90 & 7.27 & 43.02 (58) & {{{0.57}}} & {{{low state}}}\\
116403560-116403576 & 0.72 & 159.63 & 2.34 & 5.38  & 1.1 & 
2.96 &16.44 & 31.97 (41) & {{{0.68}}} & {{{low state}}}\\
116403580-116403596 & 0.68 & 177.95 & 2.56 & 9.65  & 1.18 & 
3.11 & 5.09 & 33.08 (41) &{{{0.61}}}& {{{low state}}}\\
116403600-116403616 & 0.69 & 169.5  & 2.54 & 9.52  & 1.12 & 
3.28 & 5.66 & 45.83 (41)& {{{0.57}}} &{{{low state}}}\\
116403620-116403636 & 0.70 & 164.87 & 2.64 & 12.47 & 1.03 & 
3.54 & 3.27 & 24.74 (33) & {{{0.55}}}& {{{low state}}}\\
116403640-116403656 & 0.71 & 166.15 & 2.52 & 10.64 & 0.97 & 
3.56 & 5.65 & 28.59 (36) & {{{0.54}}} & {{{low state}}}\\
116403660-116403656 & 0.67 & 178.83 & 2.72 & 15.66 & 1.12 & 
3.98 & 2.60 & 23.31 (36)&{{{0.5}}} &{{{low state}}}\\
116403680-116403696 & 0.70 & 163.07 & 2.72 & 16.51 & 1.11 & 
**** &   &  &{{{0.50}}}&{{{low state}}}\\
116403700-116403716 & 0.77 & 133.77 & 2.61 & 13.64 & 1.16 & 
4.13 & 2.77 & 49.29(45)&{{{0.51}}}&{{{low state}}}\\
116403720-116403736 & 0.77 & 126.64 & 2.67 & 17.23 & 1.05 & 
4.28 & 42.8 & 12.09 (36)& {{{0.45}}} &{{{low state}}}\\
116403740-116403756 & 0.78 & 132.98 & 2.56 & 13.47 & 0.74 & 
4.53 & 11.92 & 31.24 (40) & {{{0.53}}} & {{{low state}}}\\
116403760-116403756 & 0.79 & 123.95 & 2.71 & 20.59 & 0.96 & 
4.56 & 26.2 & 28.75 (40) & {{{0.45}}} & {{{low state}}}\\
116403780-116403776 & 0.84 & 111.20 & 2.71 & 22.32 & 0.94 & 
5.11 & 4.25 & 55.48 (46) & {{{0.46}}} &{{{low state}}}\\
116403800-116403816 & 0.93 & 85.46  & 2.77 & 27.63 & 1.16 & 
5.20 & 31.31 & 34.05 (40) & {{{0.42}}} &{{{low state}}}\\
116403820-116403816 & 0.95 & 83.65  & 2.78 & 28.53 & 0.92 & 
5.97 & 26.41 & 28.88 (40) &{{{0.44}}} & {{ low state}}\\
116403840-116403856 & 1.06 & 67.07  & 2.86 & 38.67 & 0.82 & 
6.01 & 59.82 & 21.31(50)& {{{0.41}}} & {{{high state}}}\\
116403860-116403876 & 1.5  & 33.95  & 3.68 & 83.66 & 
0.56 & 9.13 & 91.1 & 33.68 (43) & {{{0.78}}}& {{{high state}}}\\
116403880-116403896 & 1.42 & 36.22  & 3.58 & 63.32 & 0.88 & 
**** &   &  & {{{0.75}}} &{{{high state}}}\\
116403900-116403916 & 1.41 & 36.0   & 3.49 & 54.8  & 0.94 & 
**** &   &  & {{{0.73}}} &{{{high state}}}\\
116403920-116403936 & 1.37 & 38.52  & 3.41 & 42.10 & 1.10 & 
**** &   &  & {{{0.77}}} &{{{high state}}}\\ 
116409810-116409826 & 1.62 & 39.8   & 3.17 & 63.16 & 0.84 & 
**** &   &  &{{{$0.77$}}}& {{{high state}}}\\
116409830-116409846 & 1.53 & 41.36  & 2.78 & 54.84 & 0.61 & 
12.69& 52.87 & 61.67 (76)& {{{0.5}}} & {{{high state}}}\\
116409850-116409866 & 1.45 & 30.52  & 2.76 & 58.22 & 
1.21 & 7.50 & 18.21 & 60.55 (76)& {{{0.22}}} & {{{high state}}}\\
116409870-116409886 & 1.33 & 25.12  & 2.80 & 56.47 & 
1.01 & 6.41 & 16.76 & 67.47 (63) & {{{0.11}}}& {{{high state}}}\\ 
116409880-116409896 & 1.05 & 40.65  & 2.76 & 39.36 & 0.89 & 
5.61 & 10.21 & 27.24 (35)&{{{0.13}}} & {{{high state}}}\\
116409900-116409916 & 0.78 & 102.81 & 2.69 & 25.76 & 0.92 & 
4.67 & 17.85 & 53.58 (66) & {{{0.25}}} & {{{low state}}}\\
116409920-116409936 & 0.76 & 122.89 & 2.54 & 14.74 & 1.85 & 
**** &   &  &{{{0.37}}}& {{{low state}}}\\
116409940-116409956 & 0.71 & 146.78 & 2.53 & 9.85  & 0.97 & 
3.38 & 5.38 & 61.44 (76)&{{{0.49}}} &{{{low state}}}\\
116409960-116409976 & 0.73 & 143.84 & 2.46 & 7.22  & 1.29 & 
2.88 & 5.43 & 61.97 (66) & {{{0.61}}} &{{{low state}}}\\
116409980-116409996 & 0.73 & 153.16 & 2.23 & 4.08  & 1.44 & 
2.50 & 5.89 & 39.44 (66) &{{{0.68}}}&{{{low state}}}\\
116410000-116410016 & 0.72 & 143.49 & 2.18 & 6.88  & 1.00 & 
2.42 & 13.44 & 47.21 (66) &{{{0.62}}} &{{{low state}}}\\
116410020-116410036 & 0.73 & 135.01 & 2.60 & 8.8   & 1.03 & 
2.53 & 19.38 & 33.01 (35) & {{{0.60}}} & {{{low state}}}\\
116410040-116410056 & 0.76 & 131.98 & 2.47 & 6.38  & 0.98 & 
2.77 & 5.65 & 13.51 (35) &{{{0.69}}} &{{{low state}}}\\
116410060-116410076 & 0.72 & 146.71 & 2.54 & 7.6   & 1.07 & 
2.75 & 6.54 & 37.04(66) &{{{0.64}}} &{{{low state}}}\\
116410080-116410096 & 0.69 & 152.11 & 2.66 & 10.35 & 1.33 & 
**** &   &  &{{{0.57}}} &{{{low state}}}\\
116410100-116410116 & 0.68 & 180.08 & 2.58 & 9.29 & 1.32 & 
3.16 &15.5 & 29.17 (35)&{{{0.54}}} &{{{low state}}}\\
116410120-116410136 & 0.71 & 162.94 & 2.51 & 8.50 & 1.22 & 
3.64 &6.61 & 16.01 (35) &{{{0.62}}} & {{{low state}}}\\
116410140-116410156 & 0.74 & 145.68 & 2.49 & 7.55 & 1.03 & 
.70 &31.48 & 21.74 (35) & {{{0.67}}} &{{{low state}}}\\
116410160-116410176 & 0.76 & 144.19 & 2.39 & 5.97 & 0.74 & 
3.71 & 3.98 & 37.41 &{{{0.71}}} &{{{low state}}}\\
116410180-116410196 & 0.69 & 168.43 & 2.69 & 13.05& 1.54 & 
3.77 &3.92 & 28.7 (43) &{{{0.53}}} &{{{low state}}}\\
116410200-116410216 & 0.74 & 146.48 & 2.60 & 10.98& 0.98 & 
4.03 &20.15 & 29.44 (48) &{{{0.61}}} &{{{low state}}}\\
116410220-116410236 & 0.79 & 126.77 & 2.56 & 10.64& 1.25 & 
4.30 &6.61 & 32.4 (40) &{{{0.60}}} &{{{low state}}}\\
116410240-116410256 & 0.72 & 154.45 & 2.75 & 16.85& 1.15 & 
**** &  &  &{{{0.55}}}& {{{low state}}}\\
116410260-116410276 & 0.76 & 154.80 & 2.75 & 16.84& 1.16 & 
4.76 &15.88 & 33.8 (40) &{{{0.55}}} &{{{low state}}}\\
116410280-116410296 & 0.81 & 117.47 & 2.71 & 17.19& 1.10 & 
5.43 &67.5 & 32.9 (36) &{{{0.55}}} &{{{low state}}}\\
116410300-116410316 & 0.85 & 93.36  & 2.83 & 25.13& 0.96 & 
5.81 &12.91 & 26.18 (32) &{{{0.52}}} &{{{low state}}}\\
116410320-116410336 & 0.90 & 72.54  & 2.90 & 33.45& 0.85 & 
5.85 &48.75 & 38.8 (33) &{{{0.39}}} &{{ {low sate}}}\\
116410340-116410356 & 0.92 & 78.18  & 2.88 & 34.91& 0.95 & 
7.0  &36.84 & 22.3 (32) &{{{0.30}}} &{{{low state}}}\\
116410360-116410376 & 1.21 & 59.09  & 2.96 & 42.68& 0.73 & 
9.57 & 32.7 & 27.47 (36) &{{{0.34}}} &{{{high state}}}\\
116410380-116410396 & 1.43 & 38.97  & 3.56 & 69.14& 0.90 & 
**** &  &  &{{{0.61}}} &{{{high state}}}\\
\hline
\end{tabular}
\caption{Spectral and temporal fit parameters for \G1915, for
each interval.  $R^{\star}=((r_{col} /km)/(D/10kpc))^{2}\cos\theta $,
where $\theta$ is the inclination angle, and $R^{\star}$ is the XSPEC
blackbody normalization parameter.  Q is a parameter relative to the QPO
width with $Q ={\nu_{QPO}}/{\sigma_{QPO}}$.  Spectral states are defined
by the values of the disk and the power law parameters; in particular a
low state is defined by a low disk temperature ($\simlt 1keV$) and a
high radius, together with a predominance of the power law component. 
Based on the criterion of Merloni {\em et al.}, we reject data points in
the high state when the ratio ${F_{dbb}}/{F_{tot}}$ is $<.5$.}
\label{table:1915spec}
\end{table*}
\end{document}